\documentclass[iop,apjl,tighten]{emulateapj}
\usepackage{graphicx,enumitem}

\shorttitle{Tidal tails around the outer halo globular clusters Eridanus and Pal\,15}
\shortauthors{Myeong et al.}

\begin{document}

\title{Tidal tails around the outer halo globular clusters Eridanus and Palomar\,15}

\author{G.~C. Myeong$^{1,2}$, Helmut Jerjen$^1$, Dougal Mackey$^1$, and Gary S. Da Costa$^1$}
\affil{$^1$Research School of Astronomy and Astrophysics, The Australian National University,\\Canberra, ACT 2611, Australia\\
$^2$Institute of Astronomy, University of Cambridge, Madingley Road, Cambridge CB3~0HA, United Kingdom}

\begin{abstract}

We report the discovery of tidal tails around the two outer halo globular clusters, \object{Eridanus} and \object{Palomar\,15}, based on $gi$-band images obtained with DECam at the CTIO 4-m Blanco Telescope. The tidal tails are among the most remote stellar streams presently known in the Milky Way halo. Cluster members have been determined from the color-magnitude diagrams and used to establish the radial density profiles, which show, in both cases, a strong departure in the outer regions from the best-fit King profile. Spatial density maps reveal tidal tails stretching out on opposite sides of both clusters, extending over a length of $\sim$760\,pc for \object{Eridanus} and $\sim$1160\,pc for \object{Palomar\,15}. The great circle projected from the \object{Palomar\,15} tidal tails encompasses the Galactic Center, while that for \object{Eridanus} passes close to four dwarf satellite galaxies, one of which (Sculptor) is at a comparable distance to that of \object{Eridanus}.

\end{abstract}

\keywords{globular clusters: general --- globular clusters: individual (\object{Eridanus}, \object{Palomar\,15)}}

\section{Introduction}

It is widely accepted that large galaxies such as the Milky Way formed 
through accretion and merger of numerous protogalactic fragments \citep[e.g.][]{searle_1978,blumenthal_1984}. A significant fraction of the Milky Way's globular cluster population 
is thought to have been acquired by this process, and it is believed that a large portion of the current halo stellar mass may have been donated by their host dwarf galaxies via tidal disruption and mass loss \citep{mackey_2004,forbes_2010}.

There are two types of stellar structures that we expect to see around Galactic globular clusters. One can occur around accreted clusters and represents the remnant of the disrupted parent dwarf galaxy (see \citet{olszewski_2009} and \citet{kuzma_2016}). In extreme cases we might see a cluster embedded in an extended stellar stream. This is the case in the halo of M31 \citep[e.g.][]{mackey_2010} and in our own Galaxy where a number of globular clusters are associated with the disrupting Sagittarius dwarf galaxy \citep[e.g.][]{law_2010}. 

The second type consists of narrow stellar streams arising from the dynamical evolution of the cluster itself in the external tidal field of the Milky Way. Many studies have investigated the presence of tidal tails associated with globular clusters \citep[e.g.,][]{grillmair_1995,leon_2000}. It has been found that some globular clusters have stellar distributions that significantly differ from a King profile \citep{king_1962}, extending beyond the nominal tidal radius \citep{grillmair_1995,carballo_2012}.

Recently, it has been discovered that the tidal tails of some globular clusters, such as Palomar\,5 \citep{odenkirchen_2001,grillmair_2006b} and NGC\,5466 \citep{belokurov_2006,grillmair_2006}, 
extend over several hundred parsecs in physical length.
In this context it is also interesting to note that many narrow streams, like the Orphan \citep{belokurov_2006b,grillmair_2006c,grillmair_2015} and Phoenix streams \citep{balbinot_2016} exist in the inner Milky Way halo. These likely originate from completely disrupted globular 
clusters \citep{newberg_2010,martin_2014}.

The main point is that the Galactic tidal field, disk and bulge shocks, and two-body relaxation can all affect the outer structures of globular clusters in ways that we do not fully understand. Adding new examples of globular clusters with tidal tails, particularly at large Galactocentric distances, thus gives us additional information about the frequency of this phenomenon and can help to probe the outermost parts of the Galaxy. The properties of such streams can further help to constrain the dark matter distribution in the Galactic halo. For example, the gaps in the tidal tails of Palomar\,5 may tell us about the dark matter sub-halos predicted in $\Lambda$CDM cosmology \citep{ngan_2014}.

So far, most studies searching for tidal structures around Milky Way globular clusters have been restricted to relatively nearby targets \citep[e.g.,][]{grillmair_2016} with the one exception of Palomar\,14 \citep{sollima_2011}. In this Letter we report the discovery of stellar substructures around two distant halo globular clusters, \object{Eridanus} 
($R_{GC}=95.0$\,kpc) and \object{Palomar\,15} (Pal\,15; $R_{GC}=38.4$\,kpc).

\section{Data Analysis}

We used the Dark Energy Camera (DECam) at the CTIO 4-m Blanco Telescope  to obtain deep imaging around Eridanus and Pal\,15. This instrument comprises a $62$ CCD mosaic that spans a 3\,sqr\,deg field-of-view and has a pixel scale of $0.27$\arcsec \citep{flaugher_2015}.
For \object{Eridanus}, we obtained five dithered $900$s $g$-band images and $11$ dithered $600$s $i$-band images on Feb 25-27, 2014 (average seeing $\sim1.11\arcsec$). 
For Pal\,15, we obtained five dithered $360$s $g$-band images and 
five dithered $360$s $i$-band images on Sept 24-26, 2013 (average seeing $\sim1.12\,\arcsec$). 
The nights were part of the observing programs 2014A-0621 for Eridanus and 2013B-0617 for Pal\,15 -- PI: D.\,Mackey.
The raw images were preprocessed with the DECam Community Pipeline \citep{valdes_2014}, including application of the astrometric solution.
We used the resampled images (individual frames) and their corresponding weight maps for our study.

SExtractor \citep{bertin_1996} and PSFEx \citep{bertin_2011} were employed for source detection and PSF photometry. Star/galaxy separation was performed using the method described in \citet{koposov_2015}, which adopts $|SPREAD\_MODEL|<SPREADERR\_MODEL+0.003$ as the threshold for stars. The instrumental magnitudes were calibrated by crossmatching with the Pan-STARRS1 StackObjectThin catalog \citep{chambers_2016}, and then de-reddened using the \citet{schlegel_1998} dust maps with the extinction coefficients from \citet{schlafly_2011}. The inferred distances are consistent with those given in \citet{harris_1996}.

For a given target we constructed the ($(g-i)_0,g_0$) color-magnitude diagram (CMD) using all stellar objects within $r<3\arcmin$ from the nominal cluster center, and fit a fiducial line representing the locus of cluster members in the CMD.
The fiducial line for each cluster was determined empirically. The CMDs for the inner $3\arcmin$, which are dominated by cluster members, were subdivided into intervals of $g_0$ and the $3\sigma$-clipped median $(g-i)_0$ colors determined. A continuous curve through the pairs of mean $g_0$ and the corresponding median color for each interval then defines the fiducial line for the cluster members.
This fiducial line was then used to calculate a weight $w$ for each star detected across the field. The weight quantifies the likelihood of a star being a cluster member based on its distance from the fiducial line along the color axis. The weight value was calculated using the Gaussian distribution $N(x,\mu,\sigma$) centred on the fiducial line (normalized to have $w=1.0$ at the centre), with $\sigma$ corresponding to the uncertainty in $(g-i)_0$ at the $g_0$ value, which were taken from the photometric errors generated by SExtractor.

Stars within $2\sigma$ range from the fiducial line in color (that is, $w=N((g-i)_{0,*}, (g-i)_{0,fid},\sigma_{(g-i)_0})>0.135$) were considered to have a high probability of being related to the target cluster.
We will henceforth call this selection ``member stars'', although it still contains some contamination from foreground/background objects (henceforth ``background'') which happen to lie near the cluster population in CMD.
To minimise this contamination while maximising the signal due to cluster members, we further restricted our selection to an interval in $g_0$ defined to cover the region on the CMD with the greatest contrast between ``cluster'' and ``background'' stars. In addition, we ensured that the faint end of this interval was 0.5\,mag above the 50\% photometric completeness limit.

Radial density profiles and spatial density maps were generated to investigate the distribution of the member stars (Figs.\ref{f:eridanus} and \ref{f:pal15}). For the radial profile (upper left panel), we binned our field-of-view in concentric annuli with logarithmic spacing. Poisson statistics were assumed for calculating the uncertainty in each bin. The best-fitting King profile \citep{king_1962} was determined by using a python library LIMFIT \citep{newville_2014} after subtracting the background level. The background level was fixed from the total ($cluster+background$) density profile by determining the median annular density after $3\sigma$-clipping.

Two-dimensional density maps (bottom panels) were constructed by binning the DECam field into cells of size $0.33\arcmin\times0.33\arcmin$ for Eridanus and $0.59\arcmin\times0.59\arcmin$ for Pal\,15, and then smoothing with a Gaussian kernel of width $\sigma=0.8\arcmin$ for Eridanus and $\sigma=1.47\arcmin$ for Pal\,15. The bin size and sigma were chosen after testing various combinations to provide the maximum contrast for the tidal features. The median stellar density of the field (=background),  beyond 3 tidal radii from the cluster, is $5$\,star/arcmin$^{2}$ ($\sigma=2.9$\,star/arcmin$^2$) for Eridanus, $7$\,star/arcmin$^2$ ($\sigma=3.1$\,star/arcmin$^2$) for Pal\,15 before smoothing.
We generated a background map for each cluster in the same manner by using the corresponding subsample of background objects. After this map was normalised to the median $3\sigma$-clipped density of the $w>0.135$ map determined above, it was subtracted to remove any large-scale gradients or instrumental artefacts across the field.
Local RMS values were measured for each of the 62 CCDs. The median was considered to be the background RMS and used to scale the background subtracted spatial density map (bottom left panel).

\section{Eridanus}

\begin{figure*}
\includegraphics[width=180mm]{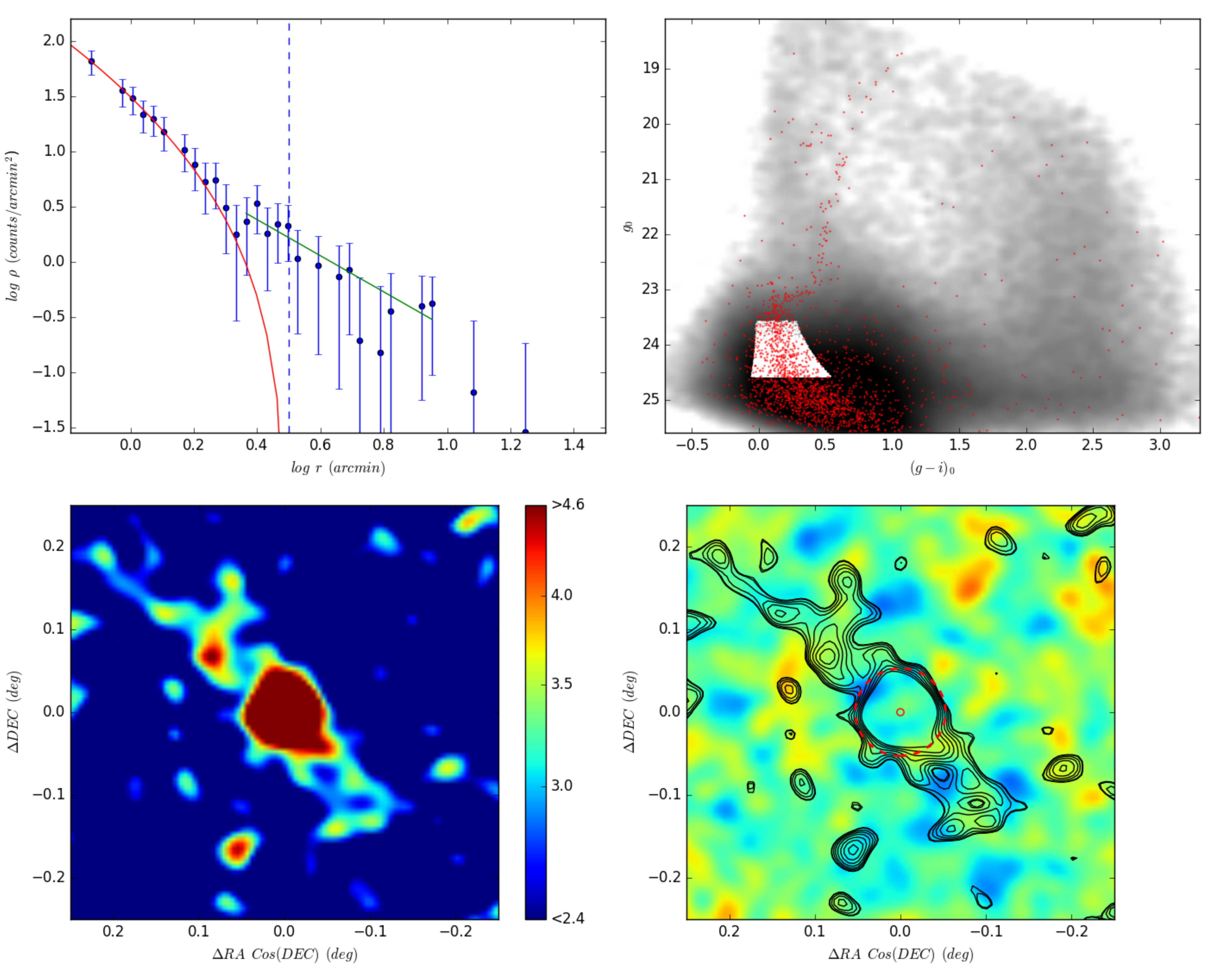}
\caption{Eridanus. {\bf Upper Left panel:} Background-subtracted radial density profile and the best-fit King profile (red line). The green line is the least-squares fit to the profile after the point where it starts to deviate from the King profile. The blue dashed line indicates the tidal radius from the best-fitting King profile. {\bf Upper Right panel:} Hess diagram of stellar objects over the DECam field-of-view. Red dots are the stars within $r<3\arcmin$ from the nominal cluster center. The white area highlights the selection box for cluster members. {\bf Lower left panel:} 2D density map of the Eridanus region after the background subtraction. The field was binned into $0.33\arcmin\times0.33\arcmin$ pixels and smoothed with a $\sigma=0\farcm8$ Gaussian kernel. {\bf Lower right panel:} Contour map generated from the lower left panel, superimposed on the background map. Circles are the core radius, $r_c=0\farcm25$ ($6.6$\,pc) (solid), and the tidal radius, $r_t=3\farcm17$ ($83.0$\,pc) (dashed) from the 
best-fitting King profile. The scale on the right side of the density map is the signal strength in units of the standard deviation above the background. The contour lines range from $2.7-4.75\sigma$ in logarithmic steps. 
\label{f:eridanus}}
\end{figure*}

Eridanus, at $R_{GC}=95.0$\,kpc \citep{harris_1996}, is one of the most distant Galactic globular clusters known. 
It has been classified as a ``young'' halo cluster, suggesting that it may have originated in an external satellite galaxy and been accreted into the Galactic halo \citep{zinn_1993,mackey_2005}. Our best-fitting King profile has a core radius $r_c=0\farcm25\pm0\farcm12$ ($6.6\pm3.1$\,pc) and a tidal 
radius $r_t=3\farcm17\pm0\farcm76$ ($83.0\pm20.0$\,pc), with the concentration index $c=\log(r_t/r_c)=1.10\pm0.23$. These values
agree well with previous measurements ($r_c=0\farcm25$ and $r_t=3\farcm15$, \citet{harris_1996}). The radial profile exhibits an excess of stars outside $r\sim1\farcm82$ $(\log(r)\sim0.26)$, 
which continues beyond the nominal tidal radius $r>r_t=3\farcm17$ and follows a power law with an index of $\gamma=-1.64\pm0.16$ (azimuthally averaged).

The spatial density map and the corresponding contour map (lower panels of Figure~\ref{f:eridanus}) reveal two prominent tidal tails extending considerably beyond the tidal radius (dashed circle in the contour map). These structures show no correlation with the background map. Tail~1 extends $\sim18\arcmin$ ($\sim480$\,pc) from the cluster center in North-East direction at a position angle of PA=$40\pm5\,^\circ$. Tail~2 extends $\sim11\arcmin$ ($\sim280$\,pc) in South-West direction (PA=$220\pm10\,^\circ$). The fractional overdensity of cluster stars is $29.4\%$ and $31.6\%$, respectively, in the distance interval $2.5\,r_t<r<3\,r_t$. The alignment of the features is close to axisymmetric suggesting that they are likely following the orbit of \object{Eridanus}. Given the alignment and the narrow width compared to the size of the cluster, it would seem likely that the features are tidal tails resulting from the dynamical evolution of \object{Eridanus} in the tidal field of the Milky Way.

\section{Palomar\,15}

\begin{figure*}
\begin{center}
\includegraphics[width=180mm]{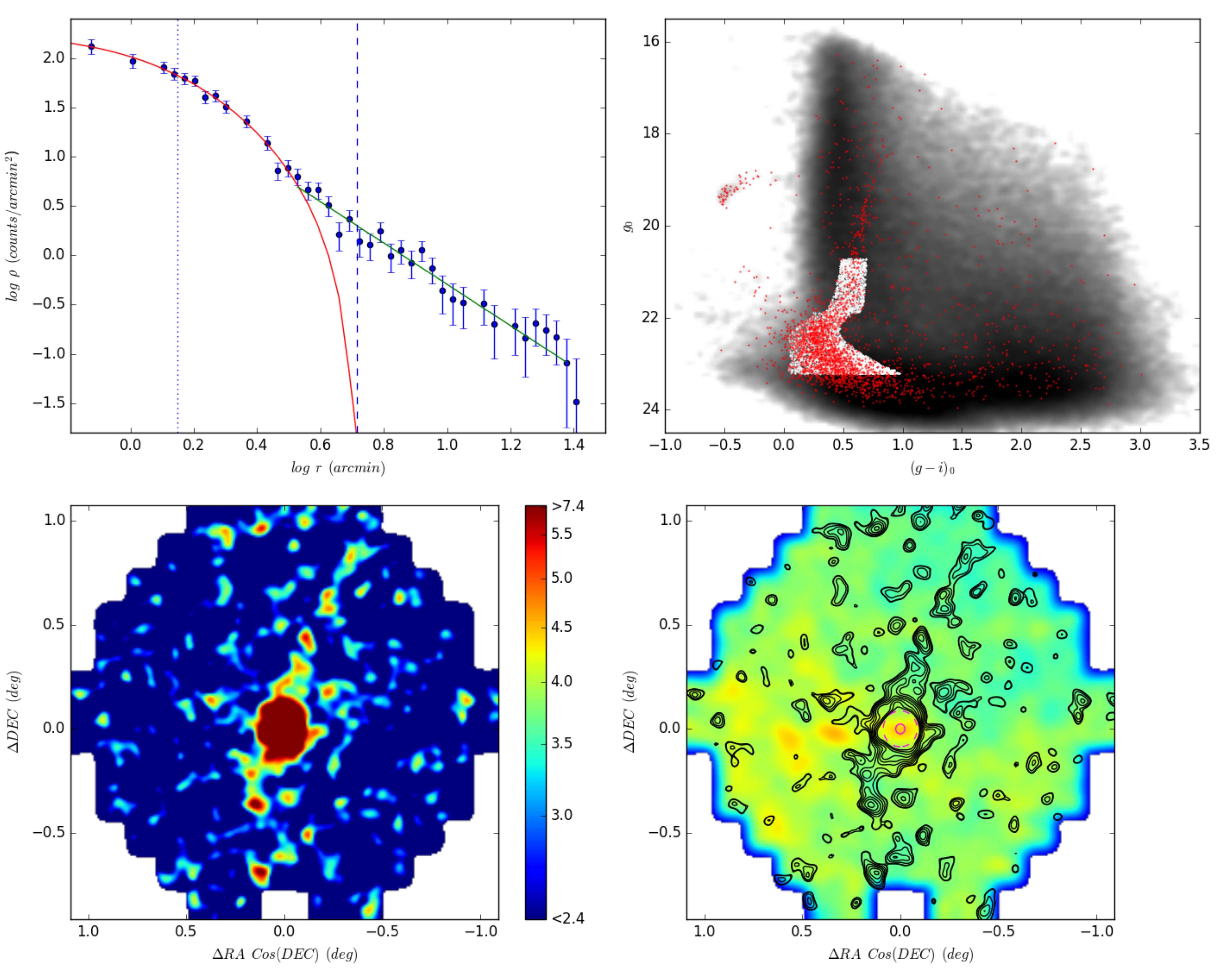}
\end{center}
\caption{The same as Figure~\ref{f:eridanus} but for Pal\,15. {\bf Upper Left panel:} The dotted line represents the core radius from the best-fitting King profile. {\bf Lower Left panel:} The field was binned into $0.59\arcmin\times0.59\arcmin$ cells and smoothed with a $\sigma=1\farcm47$ Gaussian kernel. {\bf Lower Right panel:} The two circles are the core radius, $r_c=1\farcm40$ ($18.4$\,pc) (solid) and the tidal radius, $r_t=5\farcm19$ ($68.1$\,pc) (dashed). 
The contour lines range from $3.0-8.85\sigma$ in logarithmic steps.
\label{f:pal15}}
\end{figure*}

Pal\,15 is an outer halo globular cluster at $R_{GC}=38.4$\,kpc \citep{harris_1996}. Although it is classified as an ``old'' halo cluster, Pal\,15 is a good accretion candidate by virtue of its location at a large Galactocentric distance \citep{mackey_2004}. Our best-fitting King profile has a core radius $r_c=1\farcm40\pm0\farcm15$ ($18.4\pm2.0$\,pc) and a tidal radius $r_t=5\farcm19\pm1\farcm12$ ($68.1\pm14.7$\,pc), with the concentration index $c=\log(r_t/r_c)=0.57\pm0.11$. This shows good agreement with previous measurements as listed in \citet{harris_1996} ($r_c=1\farcm20$, $r_t=4\farcm77$). The small concentration index of $0.57$, mainly a consequence of the exceptionally large core radius, suggests that Pal\,15 has been severely affected by the Galactic tidal field. Both Pal\,5 and Pal\,14 also have large core radii, and are known to have extensive tidal tails \citep{odenkirchen_2001,sollima_2011}.

The radial profile of Pal\,15 follows closely a King profile until the local star density has dropped to $\sim3\%$ of its central value
($r\sim3\farcm47$ or $\log(r)\sim0.54$), beyond which there is a strong excess of stars that continues past the nominal tidal radius and follows a power law with an index of $\gamma=-2.09\pm0.09$ (azimuthally averaged).

The spatial density and contour maps (lower panels in Figure~\ref{f:pal15}) show clear tidal tail features extending beyond the cluster's tidal radius (dashed circle in the contour map), crossing the field from the North-West to the South. These structures show no correlation with the background trend, nor with the extinction map \citep{schlegel_1998,schlafly_2011}. Tail~1 extends $\sim59\arcmin$ ($\sim780$\,pc) from the cluster centre in North-West direction at a position angle of PA=$340\pm5\,^\circ$. At the $\sim3.0\,\sigma$ significance level it consists of two main fragments (see contour map). We tested various combinations of binning sizes and smoothing kernels. The secondary, more distant fragment is visible in every solution. This supports the notion that this continuation of Tail~1 is indeed a real feature. The gap between the two segments may be similar to the gaps seen in the Pal\,5 tails, whose origin has been the subject of a number of studies \citep[e.g.][]{carlberg_2012}. Tail~2 extends $\sim29\arcmin$ ($\sim380$\,pc) in South-East direction with a possible kink at half this length. The mean position angle is PA=$150\pm10\,^\circ$. The fractional overdensity of cluster stars is $20.9\%$ and $22.5\%$, respectively, in the distance interval $2.5\,r_t<r<3\,r_t$. Both tails have considerable sizes when compared to Pal\,15's nominal tidal radius ($r_t=5\farcm19$). Although the two tails are different in length, their alignment close to axisymmetric suggests that they are following the orbit of Pal\,15.
Additionally, both tidal tails are relatively narrow compared to the cluster size, indicating that these stellar streams are, like in the case of Eridanus, the result of the dynamical evolution in the Galactic tidal field.

\section{Discussion}

\begin{figure*}
\begin{center}
\includegraphics[width=180mm]{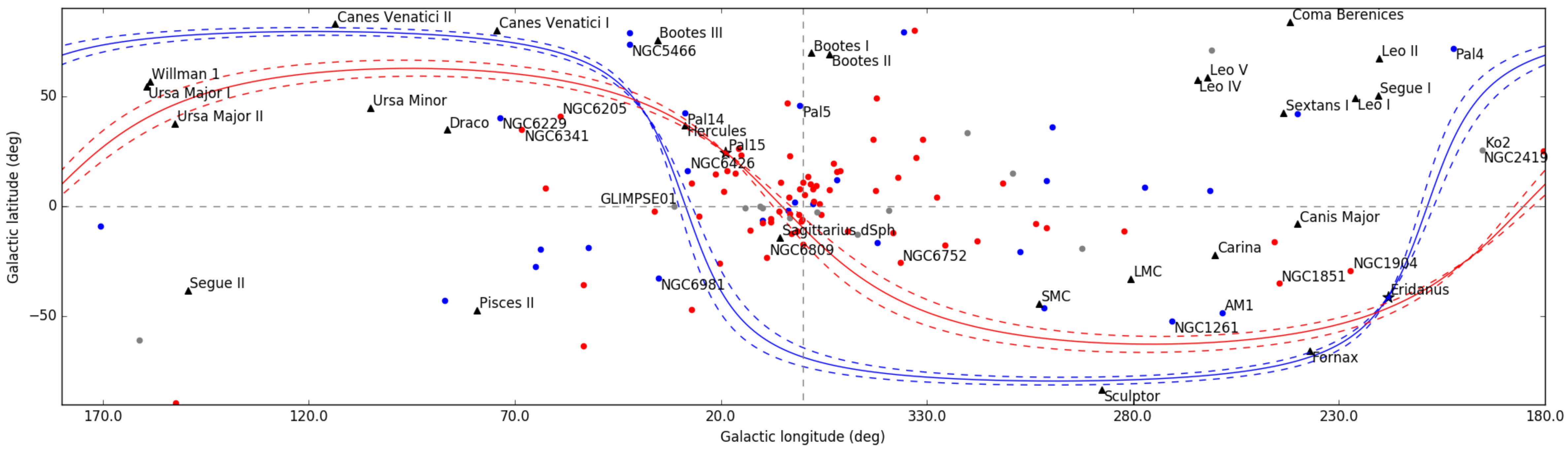}
\end{center}
\caption{Great circle along the direction of the tidal tails of Eridanus (blue line) and Pal\,15 (red line). Dashed lines in the same color indicate the corresponding uncertainty range. The two stars are Eridanus (blue) and Pal\,15 (red). Milky Way globular clusters listed in \citet{harris_1996} are marked as dots in blue for ``young'' halo clusters, red for ``old'' halo clusters, and grey for unclassified \citep{mackey_2005}. Major Milky Way dwarf satellite galaxies listed in \citet{mcconnachie_2012} are marked with black triangles. Vertical and horizontal grey dashed lines indicate $0\deg$ of Galactic longitude and latitude, respectively. \label{f:orbits}}
\end{figure*}

\begin{table*}
\begin{center}
\caption{Basic parameters for the tidal structures of Eridanus and Pal\,15 \label{tab:basic}}
\begin{tabular}{ c c c c c c c }
\hline
\hline
ID & $r_{c}$ & $r_{t}$ & $c$ & $\rm{size_{t1}}$ & $\rm{size_{t2}}$ & $\rm{angle_{inc}}$\tabularnewline
 & (pc) & (pc) & & (pc) & (pc) & (deg)\tabularnewline
\hline  
Eridanus & $6.6\pm3.1$ & $83.0\pm20.0$ & $1.10\pm0.23$ & $\sim480$ & $\sim280$ & $\sim180$\tabularnewline
Pal\,15 & $18.4\pm2.0$ & $68.1\pm14.7$ & $0.57\pm0.11$ & $\sim780$ & $\sim380$ & $\sim170$\tabularnewline
\hline
\end{tabular}
\tablecomments{($_{\rm{t1}}$) indicates Tail~1, ($_{\rm{t2}}$) indicates Tail~2, and ($\rm{angle_{inc}}$) indicates the angle between two tails.}
\par\end{center}
\end{table*}

We have discovered tidal tails extending from the two remote Milky Way globular clusters Eridanus and Pal\,15.
The narrowness of the tails compared to the cluster size, along with their symmetric orientation on either side of the cluster centres indicates that they are due to the loss of cluster members to the Galactic tidal field, as seen for several other globular clusters such as Pal\,5 and NGC\,5466, and in numerical simulations \citep{combes_1999,capuzzo_2005}. 
Moreover, the power-law slope for each tail region (PA$\pm10\,^\circ$) shows $\gamma=-1.20\pm0.19$ and $\gamma=-1.25\pm0.16$ for Eridanus, $\gamma=-1.24\pm0.12$ and $\gamma=-1.26\pm0.12$ for Pal\,15, which are similar values as in the case of Pal\,5 \citep{odenkirchen_2003} and steeper than constant density.

Table~\ref{tab:basic} contains the estimated basic parameters for the tidal tails. Both pairs exhibit significant extent beyond the nominal tidal radius of their respective cluster -- the tails of Eridanus span $\sim760$\,pc in total, while those for Pal\,15 trace $\sim1160$\,pc.

The tails of Pal\,15 may well extend beyond the edge of the DECam field-of-view. In this case, our study provides a lower limit on their length, and some additional off-field data will be required to test the true extent of Pal\,15's tidal structure. In addition, the tails appear curved, especially if the fragment near the Southern edge of our Pal\,15 field is considered to be part of the structure. If confirmed, this could hold information about the underlying Galactic potential or it may be simply related to projection effects \citep{combes_1999}. 

For Eridanus, the tidal tails are well confined to the inner part of our DECam field. This suggests that we have effectively found their full extent, unless the star densities in the tails are getting too low relative to the background. This may indicate an eccentric orbit of Eridanus about the Milky Way as we may be seeing its long tails foreshortened.

We generated the great circle along the direction of the tidal tails for each cluster, based on the position angle and its uncertainty estimated for each tail (Figure~\ref{f:orbits}). We compared its path with the positions of the Milky Way globular clusters listed in \citet{harris_1996} and the Milky Way dwarf satellites in \citet{mcconnachie_2012}. Since the tidal tails are expected to be projected along the orbital path of the cluster \citep{combes_1999, capuzzo_2005}, this great circle can provide a rough estimate for the cluster's orbit. It is notable that the potential orbit of Eridanus passes close to the dwarf galaxies Canes\,Venatici\,I, Canes\,Venatici\,II, Fornax and Sculptor. Sculptor \citep[$R_{GC}=86$\,kpc,][]{mcconnachie_2012} has a comparable distance to that of Eridanus.
Since Eridanus is classified as a ``young'' halo cluster based on its metallicity and horizontal branch morphology \citep{mackey_2005}, and indeed exhibits a CMD consistent with an age up to $\sim2$\,Gyr younger than the oldest Milky Way globular clusters \citep{stetson_1999}, this possible association to Sculptor and/or another of the three satellites might provide additional evidence of Eridanus' extragalactic origin (see also \citet{keller_2012}). Even so, it is striking that despite several indications that both Eridanus and Pal\,15 might be accreted, and our data being sensitive enough to detect low surface brightness features at their distances, no clear evidence for stellar debris from parent dwarf galaxies was observed. \citet{carballo_2015} suggested that \object{Eridanus} may be associated with the Monoceros ring based on the modelled orbit of this structure by \citet{penarrubia_2005}; however, the great circle of \object{Eridanus} disagrees with this modelled orbit, suggesting no clear association between them.

The great circle defined by the tidal tails of Pal\,15 passes close to Ursa\,Major\,II which also has a comparable distance of $R_{GC}=38$\,kpc \citep{mcconnachie_2012}, suggesting possible association.
We further notice that the potential orbits of Eridanus and Pal\,15 extend to relatively high Galactic latitudes, indicating that the clusters may be on plunging orbits relative to the Galactic disk. Assuming the orbits are sufficiently eccentric to pass through the inner region of the Milky Way, the clusters could experience periodic disk and/or bulge shocks. Since these processes are known to be destructive \citep{gnedin_1997, gnedin_1999}, this could explain why both clusters exhibit tidal tails. A prime example is Pal\,5, which suffers this type of tidal shock and is predicted to be destroyed at its next disk crossing \citep{dehnen_2004}.

In the case of Eridanus ($R_{GC}=95.0$\,kpc), we have discovered a new tidal stream in an extremely remote part of the Galaxy which is poorly understood and where no other narrow streams are known. Studying Eridanus and its tails in detail may lead us to a better understanding of the gravitational potential in the extreme outskirts of the Galaxy. A key aspect of this will be understanding whether the tails arise from the action of the tidal field at this large Galactocentric distance, or whether Eridanus is on a very eccentric orbit such that it passes through the inner Milky Way, and the tails are more likely due to the action of the tidal field at smaller radii.

\acknowledgments
The authors acknowledge the support of the Australian Research Council through Discovery
projects DP150100862 and DP150103294. This project used data obtained with the Dark Energy Camera (DECam), which was constructed by the Dark Energy Survey (DES) collaboration. Funding for the DES Projects has been provided by the DOE and NSF (USA), MISE (Spain), STFC (UK), HEFCE (UK). NCSA (UIUC), KICP (U. Chicago), CCAPP (Ohio State), MIFPA (Texas A\&M), CNPQ, FAPERJ, FINEP (Brazil), MINECO (Spain), DFG (Germany) and the collaborating institutions in the Dark Energy Survey, which are Argonne Lab, UC Santa Cruz, University of Cambridge, CIEMAT-Madrid, University of Chicago, University College London, DES-Brazil Consortium, University of Edinburgh, ETH Z\"urich, Fermilab, University of Illinois, ICE (IEEC-CSIC), IFAE Barcelona, Lawrence Berkeley Lab, LMU M\"unchen and the associated Excellence Cluster Universe, University of Michigan, NOAO, University of Nottingham, Ohio State University, University of Pennsylvania, University of Portsmouth, SLAC National Lab, Stanford University, University of Sussex, and Texas A\&M University.

\end{document}